\documentclass[twocolumn,amsmath,amssymb]{revtex4}

\usepackage{graphicx}
\usepackage{float}
\usepackage{subfigure}
\usepackage{indentfirst}
\usepackage{amsmath}
\usepackage{color}
\usepackage{mathtools}
\usepackage{multirow}
\begin{document}
\title{Role of Crown in Tree Resistance Against High Winds}
\author{Hsin-Huei Li$^1$, Yu-Chuan Cheng$^1$, Kai-Jie Yang$^2$, Chia-Ren Chu$^2$, and Tzay-Ming Hong$^{1,\ast}$}
\affiliation{$^1$Department of Physics, National Tsing Hua University, Hsinchu, Taiwan 30043, Republic of China}
\affiliation{$^2$Department of Civil Engineering, National Central University, Taoyuan, Taiwan 32001, Republic of China}

\begin{abstract}
Rather than using wooden sticks to simulate the breakage of trees  in high winds as in most research, we employed fresh samples with branches and leaves to certify the crucial role played by the tree crown.
By using the blowdown wind tunnel with a maximum wind speed of  60 m/s,
we purposely reduce the number of leaves and show that the drag force will drop by as much as two thirds when half pruned.
Based on real observations, we model the leaf by an open and full cone in the presence of light and strong wind, and calculate  how their corresponding cross-sectional area and drag force vary with wind speed. Different power-law relations are predicted and confirmed by experiments for these properties before and after the formation of a full cone. Compared to the empirical value of 1/3 and 3/4, our simple model gave  2/5 and 2/3 for the power-law exponent of cross-sectional area at low and high winds. Discrepancy can be accounted for by including further details, such as the reorientation of open cones and  the movement of branches.
\end{abstract}

\maketitle
\section{Introduction}

People have been interested at the resistance of solid matter and done many research to find out how different structures affect the strength of material. Direct applications include not only architecture, but also investigating the adaptability of plants, muscles, and bones to extreme conditions. In this study we focus on trees for which the force of twisting and bending due to wind\cite{pre} can be much harsher than gravity\cite{royal}. In recent years, climate change increases the number and severity of natural disaster\cite{pnas,frequent,natgeo}. The  resistance of trees under extreme climate has gotten attention of scientists \cite{pre}. Common wisdom tells us that wider crowns and talller trunks catch more wind during the storm. Therefore, they should bear more stress and are more likely to snap. But is it really the case in nature? Correlated by tree allometry \cite{selfsim}, the crown shape/size and trunk diameter are often associated with many factors, such as optimizing the exposure to sunlight, maximizing the accessability to soil and resistance to wind, and controlling  the loss of water through evapotranspiration \cite{light, nature, science, 2017Eloy}. 

A nice example that successfully combined experimental and theoretical studies is found in a recent work\cite{pre} by a group of French researchers. Based on data from a storm in 2009, they concluded  that all trees break down when the wind speed exceeds the critical value of roughly 42 m/s. They simulated distinct varieties and characteristics of tree by different radii of beech stick and pencil lead that were fixed horizontally at one end, while increasing the weight hung at the other end.  The radius of curvature was measured  until the sample stick broke.  By approximating trees as a cylinder, they could calculate the equivalent wind force from the curvature. Finally, they concluded that  the critical wind speed at which trees break has  nothing to do with the height, diameter, and elastic properties of trees \cite{pre}.

\begin{figure}[h]
\centering
\includegraphics[scale=0.28]{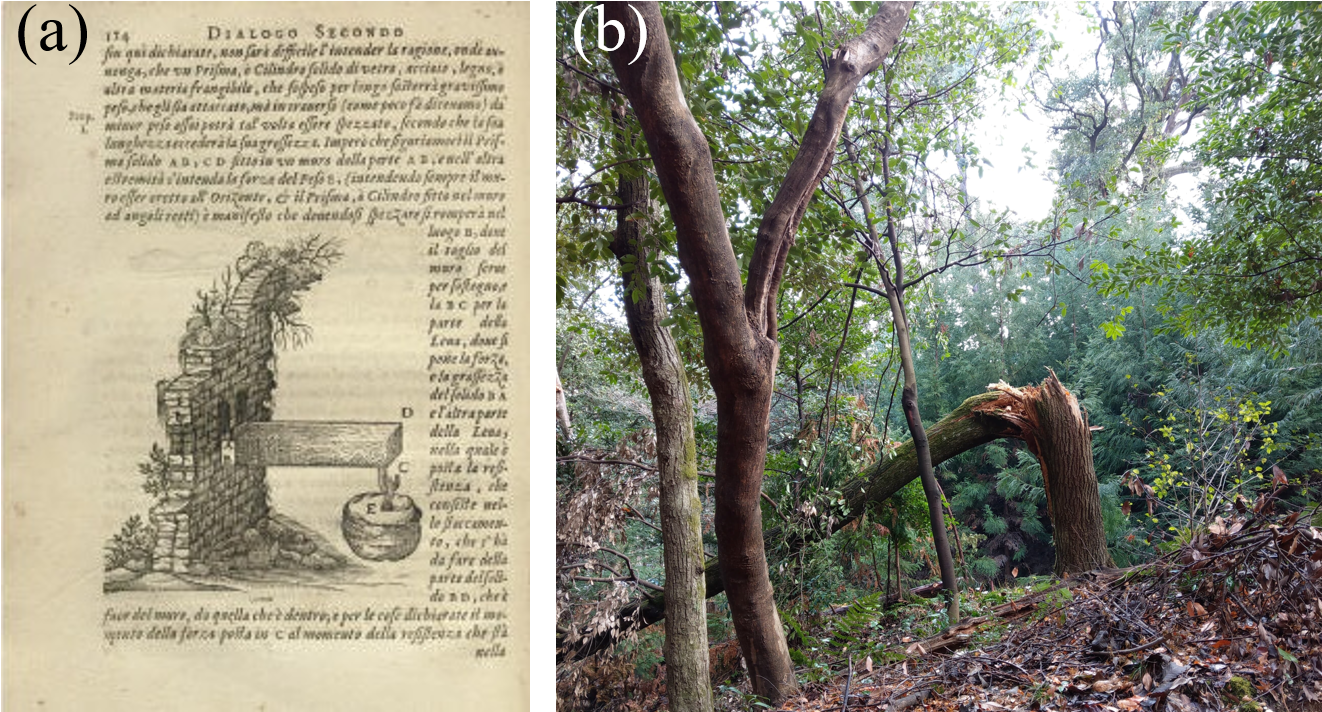}
\caption{(a) Early studies on the resistance of wood by Galileo\cite{Galilei} whose conclusion on the relation between weight and geometry of beams was disputed by Ref.\cite{Buffon}. (b) Broken trees always leave behind a considerable stump. }
\label{Fig.intro1}
\end{figure}

One interesting observation is that their experimental design \cite{pre} pretty much followed that of Galileo \cite{Galilei} from the 17th century in Fig. \ref{Fig.intro1}(a). This tradition in fact can be traced further back to  Da Vinci \cite{da} and has been  observed by later researchers \cite{Buffon}. 
Although the conclusions of Ref. \cite{pre}  seemed to match the field observations, there are important loose ends to be tied \cite{Comment}. For instance, their pencil lead always breaks at the fixed end which is apparently unlike the fate of real trees in Fig. \ref{Fig.intro1}(b). Aside from details like this, a more pressing question is how realistic it is to simulate a tree by a cylinder. Most arbonists and urban foresters will protest that branches and leaves ought to play a major role when discussing tree resistance to wind. Otherwise, cutting and pruning of trees will not be regarded as an effective precautionary measure for high winds. Academically a sensible question to ask is how the tree crown contributes to the drag force, while at the same time its cross-sectional area $A$ depends  sensitively on the wind speed $v$. To obtain $A(v)$, it is thus necessary to obtain intimate knowledge on how  leaves are distorted under different magnitude of wind. 

There have been many wind tunnel experiments \cite{1973Mayhead} in countries where forest industries were well developed since the last century. Among them, a series of noticeable papers \cite{2004Rudnicki,2005Rudnicki} by Mitchell {\it et al.} were dedicated to studying several species of conifers and hardwoods, that concluded (1) the relationship between drag force $F$ and the product of branch mass and $v$ is near-linear, and (2)  the qualitative pattern for how the frontal area of trees was reduced by increasing wind speed was not substantially changed by pruning. Unfortunately, the speed of their  wind tunnel was limited to under 20 m/s.

A useful survey on $F(v)$ was made by Cullen\cite{2005Cullen} who organized literatures that have reported or commented on a linear increase in wind drag by trees with wind speed, and tried to explore possible explanations behind their relationship. Although the data available to him was similarly limited to $v\le 20$ m/s, he concluded that $F\propto v^2$ at high $v$ is not inappropriate.

In the mean time, there were other  researchers who paid attention to the behavior of leaves under high winds. For instance, Vogel \cite{1989Vogel} indicated that some broad leaves reconfigure into increasingly acute cones and their drag coefficients decrease with increasing $v$, but he only provided data for $10<v\le 20$ m/s when the leaves already form a full cone. Consider the urgency to understand the role of  crown in tree resistance under  severe hurricanes \cite{saffir} whose frequency has increased in recent years, it is imperative to revisit these studies for a wider range of $v$.

\section{Experimental Setup}

\begin{figure}[h]
\centering
\includegraphics[scale=0.38]{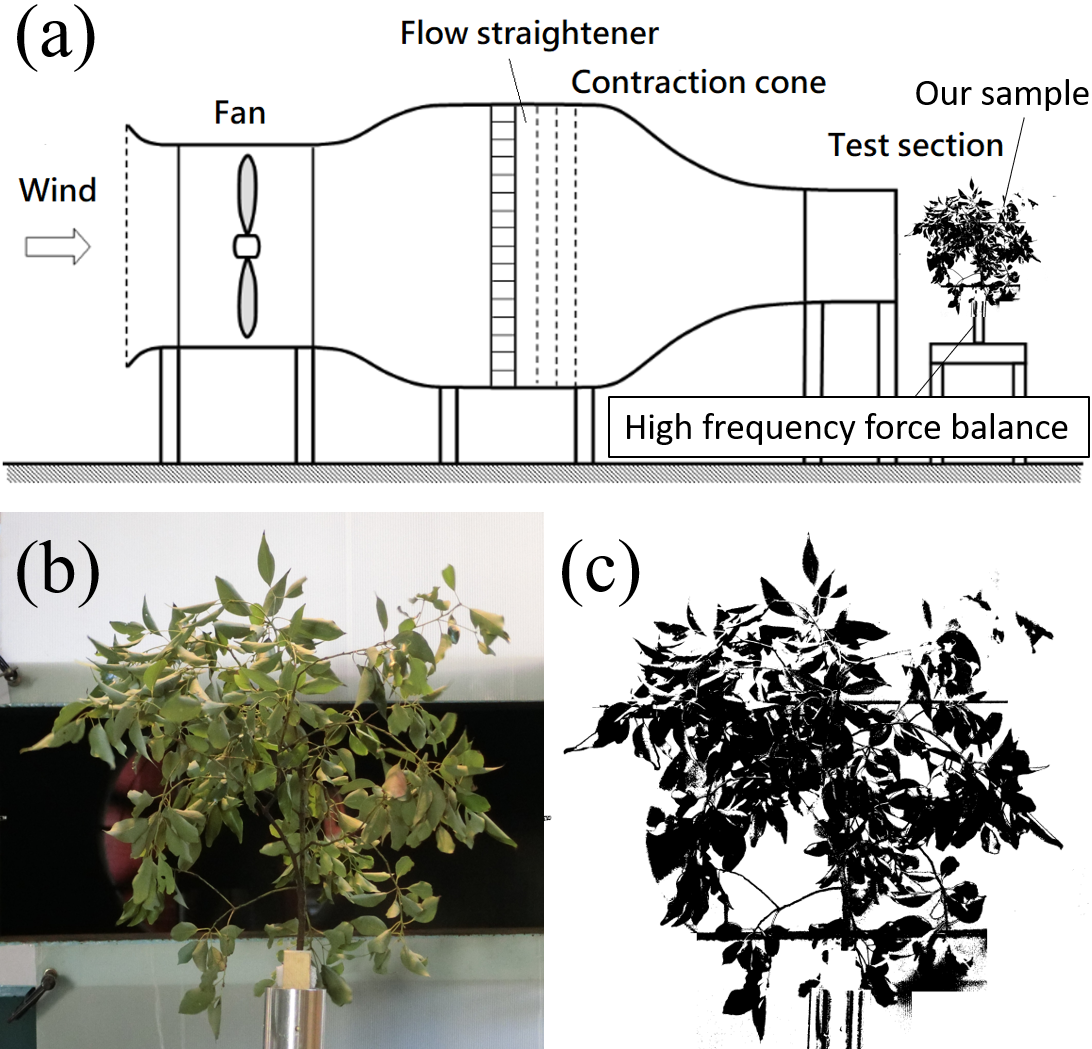}
\caption{(Color online) (a) Branch with fresh leaves is attached to a high-frequency force balance on the right side of the schematic structure. Picture in (b) is taken from the leeward  at zero wind speed. To facilitate the determination of cross-sectional area, photo (b) is  processed into the silhouette image in (c).  }
\label{Fig.main1}
\end{figure}

We use the blowdown wind tunnel in Fig. \ref{Fig.main1}(a) to simulate the blowing wind.  It can be divided into four parts: fan, flow straightener, contraction cone, and test section. The test section has a height of 0.5 m and a width of 1.2 m with a maximum wind speed of  60 m/s,  equivalent to strong typhoons or level 17 in the Beaufort scale \cite{saffir}.

Our sample mainly consists of {\it Cinnamomum camphora}, commonly known as camphor tree, that is native to and easily accessible in Taiwan. Fresh sample comprising both branches and leaves is retrieved and connected to a high-frequency force balance to measure the magnitude of wind force experienced by the sample. There are five pairs of strain gauges in the  force balance.  When the wind starts to pound on the sample, a voltage change can be picked up and transformed to force. Note that the force balance need to be calibrated before the measurement so that the estimated force falls in the valid range of the balance.

We also take pictures of the sample from leeward, such as Fig. \ref{Fig.main1}(b), to compute $A$ under different $v$. With the aid of image processing \cite{imagej}, we distinguish the actual area occupied by the sample from the background, as in Fig. \ref{Fig.main1}(c). This information is useful not only because it quantifies the portion of tree crown that takes the brunt of the wind, but also verifies that leaves and branches and boughs are far from being passive in their interactations with the wind. It is interesting to study how and why they adjust themselves under different $v$. 

\section{Experimental Results}

\begin{figure*}[ht]
\centering
\includegraphics[scale=.38]{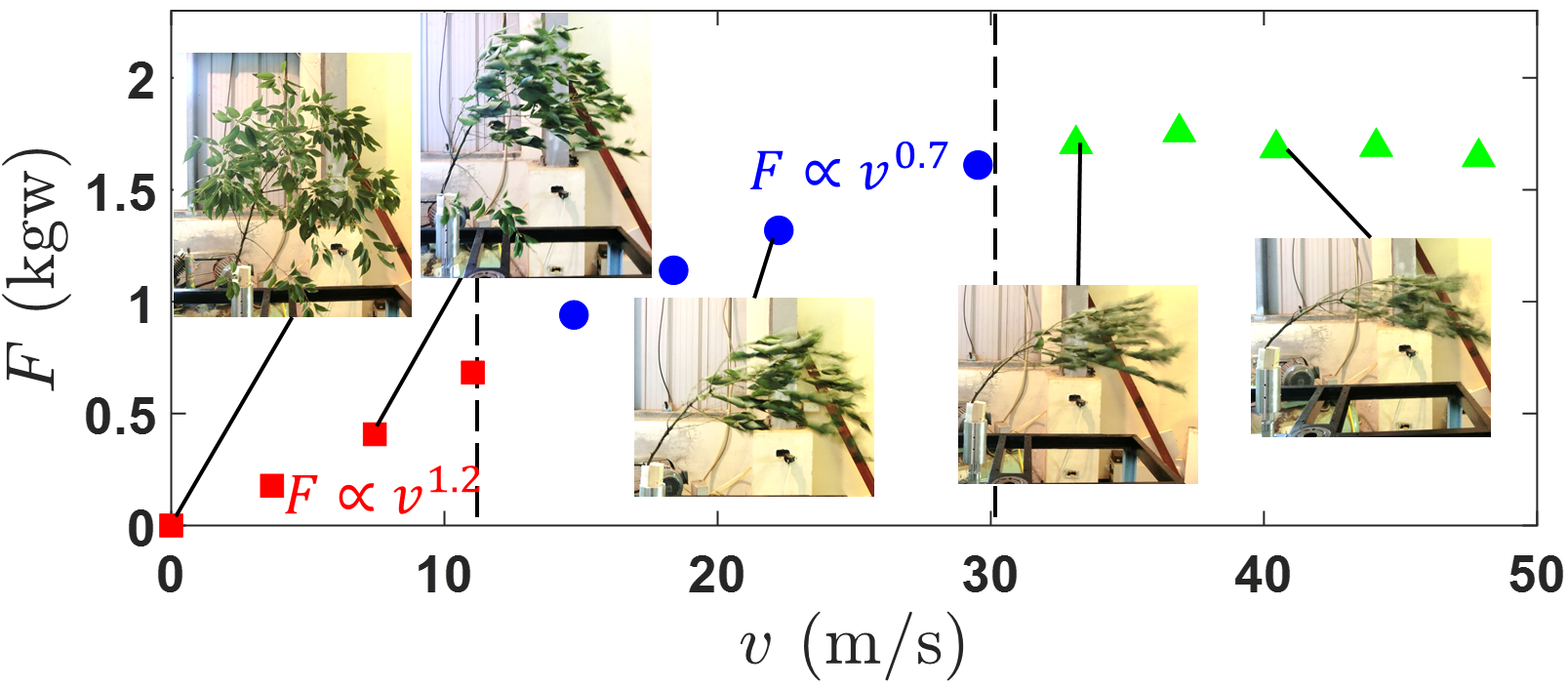}
\caption{(Color online) $F$ vs. $v$. Photos convey the degree of distortion to leaves. The two vertical dash lines separate three regimes where leaves transfigure from an open to full cone at $v\sim$ 11 m/s and become shattered as $F(v)$ levels off beyond $v\sim 30$ m/s. Although it may not be statistically meaningful due to the limited range, two power-law fittings are provided for practical purposes.}
\label{Fig.main3}
\end{figure*}

As expected, the force $F(v)$ experienced by the sample increases with  $v$ in  Fig. \ref{Fig.main3} and can be divided into three regimes at around 11 and  30 m/s. When $v$ is low, $F$ is roughly  linear, consistent with the previous research \cite{1973Mayhead,2004Rudnicki,2005Rudnicki,2005Cullen}. What has not been reported is that $F$ eventually levels off  beyond $v\sim$30 m/s. We ascribe this saturation to the observation that leaves start to be torn and blown off by the high winds, which compensates the increment of wind force experienced by the remaining leaves. Note that this implies the leaves play a dominant role when a tree tries to adjust itself to minimize the damage brought by the adverse environment, as opposed to the relatively passive branches, boughs, and trunk.

%\onecolumngrid

%\twocolumngrid

\begin{figure}[b]
\centering
\includegraphics[scale=0.25]{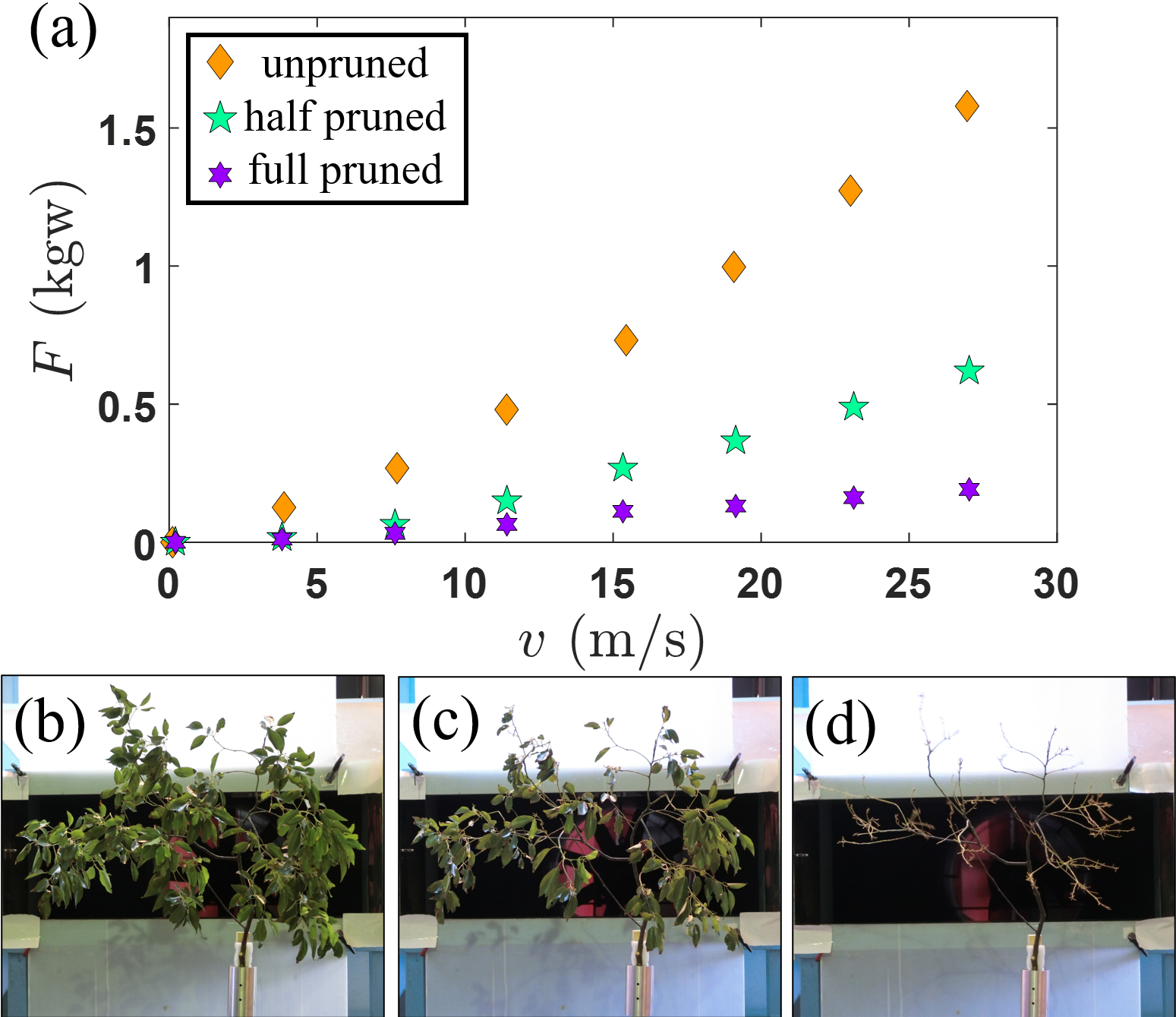}
\caption{(Color online) (a) Pruning is shown to have a significant effect on $F(v)$. Diamonds represent unpruned branches, while pentagrams and hexagrams denote half and full pruning. Their corresponding photos are shown in (b$\sim$d). }
\label{Fig.main4}
\end{figure}

Next, we want to investigate how the density of leaves influences the resistance of trees. In contrast to the original sample, we arrange to trim half of the leaves and the whole leaves in two separate experiments. As shown in Fig. \ref{Fig.main4}, the extent of pruning has a noticable effect on $F(v)$.  This is a clear evidence that the crown, leaves in particular, must be taken into consideration when estimating the air drag or the critical wind speed at which trees break \cite{pre}.

\begin{figure}[b]
\centering
\includegraphics[scale=0.30]{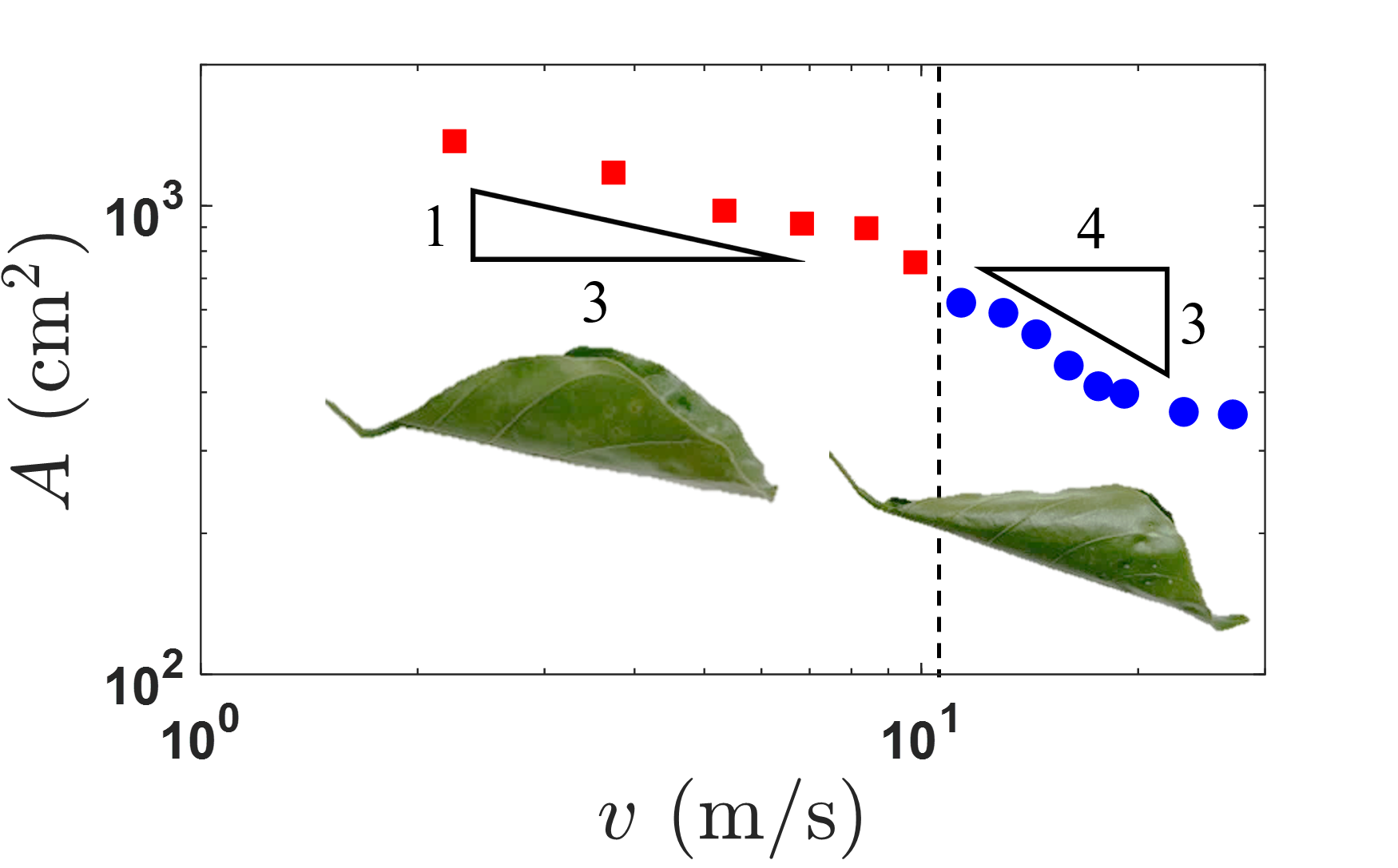}
\caption{(Color online) Decay of the cross-sectional area $A$ with $v$ in full-log scale. The vertical dash line marks the simultaneous change of power-law exponent and leaf configuration. Although the range of validity is less than one order, the fitting is to enable comparison with the theoretical prediction.}
\label{Fig.main5}
\end{figure}

Being an important parameter in our problem, the cross-sectional area of our sample is plotted against wind speed in Fig. \ref{Fig.main5} which is indicative of power-law behavior, $A\propto v^{-\alpha}$. Coincide with the configurational change at $v\sim 11$ m/s in Fig. \ref{Fig.main3}, the empirical value of $\alpha$ is found to transit from 1/3 to 3/4. This  implies that the tree crown undergoes a more dramatic rate of reduction in cross section at high winds, although the leaves have already been rolled up in the shape of a full cone.

\section{Theoretical Model}
\begin{figure}[h]
\centering
\includegraphics[scale=0.20]{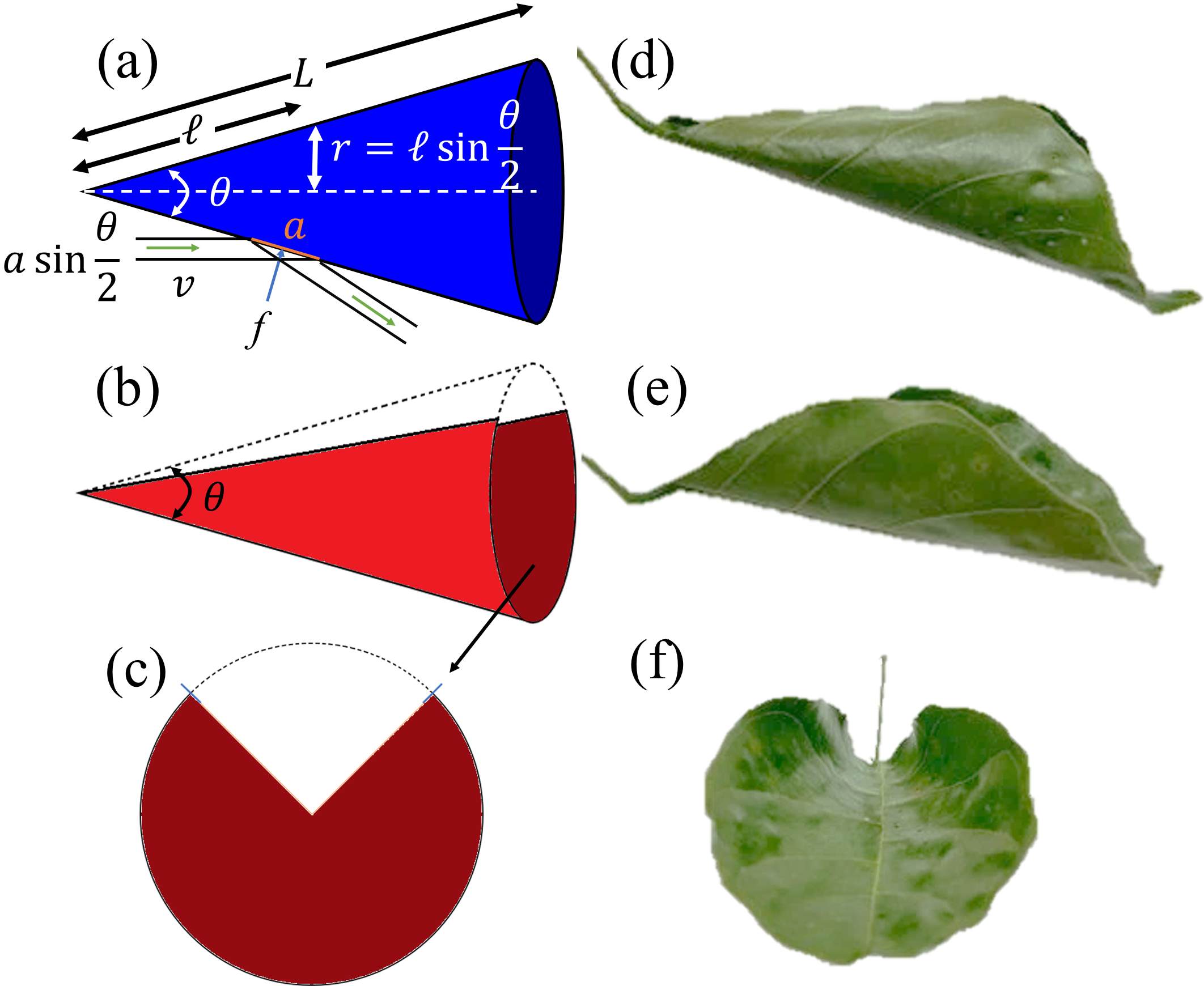}
\caption{(Color online) (a) The leaf is rolled into a full cone with apex angle $\theta$. The impinging wind molecules exert a normal force $f$ on the tiny area $a$. (b, c) are the side view and cross section of an open cone.  (d$\sim$f) show real leaves exemplified by (a$\sim$c).}
\label{Fig.main6}
\end{figure}

Let's simplify the  leaf shape as being fanlike,  for convenience. The radius of sector $L$ represents the leaf length, and the central angle $\phi$ multiplied by $L$ acts for the width. When blown by the wind, leaves roll into a cone \cite{1989Vogel} with an apex angle $\theta$ and height $L \cos (\theta/2)$, as indicated in Fig. \ref{Fig.main6}(a).  The radius of curvature at a distance $\ell$ from the apex equals $r=\ell \sin (\theta/2)$. By integrating over different slices of $\ell$, the total elastic energy $V$  of the rolled-up leaf can be calculated as
\begin{equation}
V=\frac{k_B}{2} \frac{1}{\sin^{2} (\theta/2)} \int_{0}^{L} \int_{0}^{\phi} \frac{1}{\ell}\ d\varphi\ d\ell
\end{equation}
where $k_B$ denotes the bending modulus and the spurious divergence at $\ell=0$ can be neglected. Since $\theta$ is mostly much less than 1, differentiating $V$ with respect to $\theta$ renders a restoring torque
\begin{equation}
\tau \propto \theta^{-3}
\label{rtau}
\end{equation}

In the mean time, we can compute the normal force per unit surface area experienced by the leaf due to the constant pounding by wind molecules as $2\rho v^2 \sin^2 (\theta/2)$ where $\rho$ denotes the air density and elastic collisions are assumed without loss of generality. Multiply it by the moment arm and sum over the cone area gives the external torque that tries to roll up the leaf:
\begin{equation}
\tau=4 \pi \rho v^2 \sin^3 (\theta/2) \int_{0}^{L} \ell^2 d\ell.
\label{tau}
\end{equation}
Equilibrate Eq. (\ref{tau}) and Eq. (\ref{rtau}) immediately gives $\theta \propto v^{-\frac{1}{3}}$ and $A \propto v^{-\frac{2}{3}}$ since $A=\pi L^2 \sin^2(\theta/2)$.

We can also compute the drag force  on  the leaf:
\begin{equation}
\begin{split}
F&=\int_{A} 2\rho v^2 \sin^2 (\theta/2) \sin (\theta/2) da
\end{split}
\end{equation}
where $v$ and $\sin (\theta/2)$ do not vary with the surface area. Counting the  extra $\sin (\theta/2)$ introduced by the surface integral, $F \propto v^2 \theta^4 \propto v^{\frac{2}{3}}$ - close to but slightly different from the linear dependence predicted by previous researchers\cite{1973Mayhead,2004Rudnicki,2005Rudnicki,2005Cullen}.

For $v$ less than 11 m/s, Fig. \ref{Fig.main3} already suggested that the impulse from wind is not enough to fully roll up the leaf and what ends up is an open cone, as shown schematically in Fig. \ref{Fig.main6}(b). In this case, the restoring torque  is the same as Eq. (\ref{rtau}) for a full cone because the elastic energy $V$ still integrates over the whole leaf surface. However, the external torque will be diminished and need to be recalculated. Replace $da$ by $\ell\ d\ell\ d\varphi$ and get
\begin{equation}
\tau=2 \phi \rho v^2 \sin^2 (\theta/2) \int_{0}^{L} \ell^2 d\ell
\label{tauh}
\end{equation}
which implies $\tau \propto v^2 \theta^2$, as opposed to $\tau \propto v^2 \theta^3$ in Eq. (\ref{tau}) for a full cone. Set Eq. (\ref{tauh}) equal to Eq. (\ref{rtau}), we get $\theta \propto v^{-\frac{2}{5}}$. Different from the full cone case, $A$ only covers a partial circle and is easy to determine as  $A \propto v^{-\frac{2}{5}}$.

How does the opening on the cone affect the drag force? Equally straightforward calculations give
\begin{equation}
F=2 \phi \rho v^2 \sin^3 (\theta/2) \int_{0}^{L} \ell\ d\ell
\end{equation}
that can be simplified to $F \propto v^{\frac{4}{5}}$, consistent with Fig. \ref{Fig.main3}.

\section{Conclusions and Discussions}
By extending the upper bound of wind speed from 20 to over 50 m/s, we found that the drag force remains roughly proportional to $v$ before saturating at around 30 m/s when more and more leaves   cease to contribute since they are either shattered or blown off by the high wind. Rather than using wooden sticks to simulate trees, we employed fresh camphor samples with branches and leaves. In order to prove that the tree crown plays a crucial role in the resistance in trees to high winds, we purposely reduced the number of leaves and shows that the drag force diminishes sensitively with pruning.

The leaves are found to be rolled up by the wind. As $v$ intensifies, the two margins gradually approach and close on each other.
Based on this observation, we model the leaf by an open and full cone at low and high winds, and calculate the $v$-dependence of their corresponding cross-sectional area $A\propto v^{-\alpha}$ and drag force. As opposed to the empirical value of 1/3 and 3/4,  our model predicts $\alpha$=2/5 and 2/3 for low and high winds. Considering the simplicity of our model, its ability to capture the right ballpark figure and a larger value at large $v$ is impressive. The discrepancy can be remedied by including more details, such as (1) the gap on  the open cone will cause an imbalance in normal force and cause the right side of cone in Fig. \ref{Fig.main6}(b) to tilt upwards. Roughly this will diminish $A$ by multiplying it by cosine of the tilt angle. And a smaller $v$ implies a larger gap, a bigger tilt, and more decrement for $A$. This correction will cause 2/5 to decrease and move closer to 1/3. (2) After most leaves have formed a full cone beyond $v>11$ m/s, the next heirarchical structure to deform should be the branches. We expect them to be drawn closer to each other by the high wind. In hindsight, this appears to drive some cones out of hiding for some hydrodynamic reasons that we do not yet fully comprehend.  

\begin{figure}[h]
\centering
\includegraphics[scale=0.28]{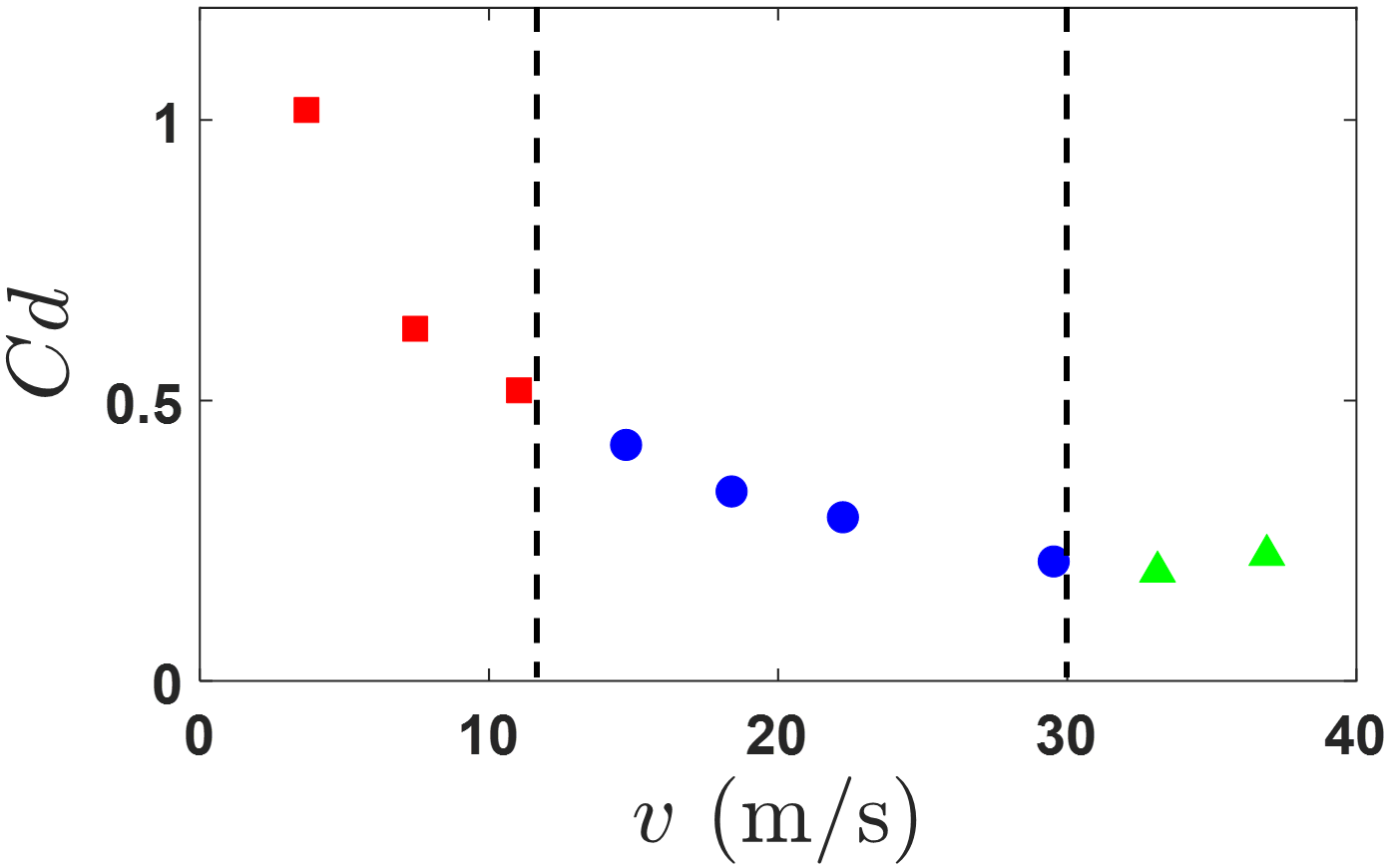}
\caption{(Color online) Drag coefficient $C_d$ vs. $v$ for samples from camphor tree.}
\label{Fig.main7}
\end{figure}

Previous researchers\cite{1973Mayhead, 2004Rudnicki, 2005Rudnicki, 2005Cullen} also concerned themselves with how the drag coefficient $C_d$ of trees varies with $v$. Our data  in Fig. \ref{Fig.main7} put $C_d$ in the same range [0.2, 1] set by previous research\cite{1973Mayhead, 2004Rudnicki, 2005Rudnicki, 2005Cullen} where $C_d$ decreases as $v$ intensifies. Virot {\it et al.} \cite{pre}  used $C_d \approx 1.0$ to compute the critical wind speed at which trees break - presumably in the high-wind limit. Judged by Fig. \ref{Fig.main7}, this is  an overestimation. If modify $C_d$ to a more realistic value of 0.2, their calculated value for the critical wind speed will jump from 56 to 125 m/s, further deviate  from the destined value of 42 m/s. 

\section{Acknowledgement}
We gratefully acknowledge technical assistance from Sheng-Han Hsieh, Li-Jie Chiu, and Professor Wei-Keng Lin, and financial support from MoST in Taiwan under Grants No. 105-2112-M007-008-MY3 and No. 108-2112-M007-011-MY3.

\end{document}